\documentclass[5p]{elsarticle}

\usepackage{hyperref}

\journal{Parallel Computing}

\begin{document}

\begin{frontmatter}

\title{SAGE: Percipient Storage for Exascale Data Centric Computing}

\author{Sai Narasimhamurthy\corref{mycorrespondingauthor}, Nikita Danilov, Sining Wu, Ganesan Umanesan}
\address{Seagate Systems UK, UK}

\author{Stefano Markidis, Sergio Rivas-Gomez, Ivy Bo Peng, Erwin Laure}
\address{KTH Royal Institute of Technology, Sweden}

\author{Dirk Pleiter}
\address{J\"ulich Supercomputing Center, Germany}

\author{Shaun de Witt}
\address{Culham Center for Fusion Energy, UK}


%
%
%

\begin{abstract}
We aim to implement a Big Data/Extreme Computing (BDEC) capable system infrastructure as we head towards the era of Exascale computing - termed SAGE (Percipient \textbf{S}tor\textbf{AG}e for \textbf{E}xascale Data Centric Computing). The SAGE system will be capable of storing and processing immense volumes of data at the Exascale regime, and provide the capability for Exascale class applications to use such a storage infrastructure. 

SAGE addresses the increasing overlaps between Big Data Analysis and HPC in an era of next-generation data centric computing that has developed due to the proliferation of massive data sources, such as large, dispersed scientific instruments and sensors, whose data needs to be processed, analyzed and integrated into simulations to derive scientific and innovative insights. Indeed, Exascale I/O, as a problem that has not been sufficiently dealt with for simulation codes, is appropriately addressed by the SAGE platform.

The objective of this paper is to discuss the software architecture of the SAGE system and look at early results we have obtained employing some of its key methodologies, as the system continues to evolve.

\end{abstract}

\begin{keyword}
SAGE Architecture \sep Object storage \sep Mero \sep Clovis \sep PGAS I/O \sep MPI I/O \sep MPI Streams 
\end{keyword}

\end{frontmatter}

\section{Introduction}
Exascale computing is typically characterized by the availability of infrastructure to support computational capability in the order of an ExaFLOP. This definition is now more broadly understood to include the storage and processing of an order of an Exabyte of data as part of a scientific workflow or a simulation. Based on various international Exascale roadmaps~\cite{dongarra2011international}, we envision Exascale computing capable infrastructures, capable of exploitation by applications and workflows for science and technological innovation, to be available to the community sometime in the 2021-2023 timeframe. 

Computing infrastructure innovation has been driven by Moore's law and the development of even more parallelism with multi-core and many-core processing to accommodate the increasing performance requirements of Exascale class problems. As an example, compute core concurrencies (billion-way concurrency on some machines!) at Exascale will have increased about 4,000 times compared to early PetaFLOP machines. However I/O and storage have lagged far behind computing. Storage performance in the same time period is predicted to improve only 100 times, according to early estimates provided by Vetter et al.~\cite{vetter2009hpc}. In fact, from the time of publication of this work, the performance of disk drives per unit capacity is actually decreasing with new very high capacity disk drives on the horizon. Simultaneously, the landscape for storage is changing with the emergence of new storage device technologies, such as flash (available today) and the promise of non-volatile memory technologies available in the near future. The optimal use of these devices (starting with flash) in the I/O hierarchy, combined with existing disk technology, is just beginning to be explored in HPC~\cite{FastForward} with burst buffers\cite{liu2012role}. 

The SAGE system ("SAGE") proposes hardware, to support a multi-tiered I/O hierarchy and associated intelligent management software, to provide a demonstrable path towards Exascale. Further, SAGE proposes a radical approach in extreme scale HPC, by moving traditional computations, typically done in the compute cluster, to the storage system, which provides the potential of significantly reducing the energy footprint of the overall system~\cite{reed2015exascale}. This design helps to move towards the Performance/Watt goals of Exascale class systems~\cite{subcommittee2014top}.

The primary objective of this paper is to present the architecture of the SAGE software stack ("SAGE stack") providing more detail on our introductory work~\cite{narasimhamurthy2017storage}. We also present early results from \textbf{two} of the components of the SAGE stack specifically in the area of programming models, which are part of High Level HPC APIs in Figure~\ref{fig:Figure2} clearly recognizing that results for the various other stack components is something we are planning to present in upcoming works as a follow-up to this work. The paper is organized as follows. Section~\ref{sec-challenges} describes the challenges that storage systems need to address to support scalable and high performance I/O on an Exascale supercomputer, and it also introduces how the SAGE project meets these challenges. Section~\ref{sec-architecture} describes the SAGE platform architecture and software stack. Section~\ref{sec-results} presents the selected results of the components of the SAGE stack. Section~\ref{sec-relwork} describes the related work. Finally, Section~\ref{sec-conclusions} summarizes the paper and outlines the future work.

\section{Exascale Challenges for Storage Systems} 
\label{sec-challenges} 
Our goal was to design and develop an I/O system with an associated software stack to meet the challenges Exascale computing poses. We identified five exascale challenges together with possible approaches to meet these challenges as part of building the SAGE system.

\textbf{1. Heterogenous Storage Systems.} The ability to store extreme volumes of data and manage exceptional data rates requires storage device technologies that are capable of meeting a wide spectrum of performance and capacity points. This capability is only made possible by having an architecture that can reasonably incorporate a wide variety of storage device technologies to meet various performance and capacity objectives at reasonable costs, as there is clearly no single storage device type that works optimally for all workloads. Further, most workloads require a variety of storage types to deliver good performance at an acceptable cost \cite{schulte2015achieving}. Indeed, this requirement is well recognized by the European as well as the International community. Storage systems at Exascale need to incorporate a deep I/O hierarchy consisting of various device types \cite{BDEC, lavignonetp4hpc}. In the SAGE platform, the top tiers consist of NVRAM pools that have higher performance but lower capacity, which hosts pre-fetched data, absorb I/O bursts, and then drain to lower tier devices, such as disks and archive, which are optimized for capacity.  To address the difficulty of optimized data placement on different tiers of the SAGE system \cite{peng2016exploring}, we further developed a series of tools which includes \textit{Hierarchical Storage Management} ('HSM'), \textit{ARM Forge} for characterizing I/O workloads and the \textit{RTMHS tool }to recommend data placement on heterogenous systems \cite{peng2017rthms}.

\textbf{2. Performance at Scale.} Existing POSIX I/O semantics cause severe performance and scalability bottlenecks that were recognized and became very well understood following many efforts to scale out HPC \cite{lang2009performance}. This behavior is a result of POSIX's very strict transactional and consistency semantics - which makes it unsuitable for highly shared environments, such as those in scale-out HPC. Further, much of the POSIX metadata needed to meet its semantic requirements are not interesting for many applications. There is a clear need to move beyond POSIX semantics and adopt new I/O models. 

Object storage technology, which has been successful in meeting the extreme volume constraints in the cloud community, is already being investigated within HPC and we aim to utilize this approach as a software framework for the platform \cite{chandrasekarexploration}. Within SAGE, we use object storage technology and in particular Mero which is Seagate's object storage platform is utilized as a base, with an API that is called Clovis. An Object Storage API is required that can ingest data from external data sets as well as handle data from simulations. The Object Storage API will need to provide a seamless view of data, independent of its location in the I/O hierarchy. Most importantly, it allows to scale the performance avoiding POSIX I/O limitations.

\textbf{3. Minimize Data Movement.} We need a system that can provide compute capability within storage to run various data analytics and pre/post processing pieces of the workflow in parallel with running simulations. The ability to run in-storage compute is crucial to avoid the energy costs incurred by constantly moving data between the compute and the storage system \cite{piernas2007evaluation}. Within SAGE, we use the function-shipping capability of the Mero object storage platform to compute in-storage and use MPI streams for offloading computations to processes running on I/O servers \cite{peng2017preparing}.

\textbf{4. Availability and Data Integrity.} Failures in infrastructure and applications are not an exception, but rather the norm at Exascale \cite{snir2014addressing}. The sheer number of software and hardware components present in Exascale systems makes the likelihood of failure extremely high during a short interval of time. A highly available infrastructure is needed that is always-on from an application perspective. The sheer volume of data in Exascale storage systems make the probabilities of data corruptions exceedingly high. Within SAGE, we provide data integrity and availability via the Mero object storage platform.

\textbf{5. Legacy and Emerging HPC Applications.} The platform should support appropriate use case data formats and legacy application interfaces (parallel file systems, POSIX) to enable their smooth transition to Exascale. SAGE also needs to interface with emerging big data analytics applications (on top of the API) to access the rich features of these tools, and the Volumes, Velocity and Variety (potentially) of data coming from sensors, instruments and simulations. We studied a portfolio of scientific data-centric applications that have been used to provide requirements to the development of the SAGE system and to validate the developments in the projects. Th applications ("SAGE applications") we chose included:
\begin{itemize}
\item  iPIC3D is a massively parallel Particle-in-Cell Code for space-weather \cite{markidis2010multi, peng2015formation}.
\item  NEST is simulator for spiking neural network models in brain science \cite{gewaltig2007nest}.
\item  Ray is a massively parallel meta-genome assembler \cite{boisvert2012ray}.
\item  JURASSIC is a fast radiative transfer model simulation code for the mid-infrared spectral region \cite{griessbach2013scattering}.
\item  EFIT++ is a plasma equilibrium fitting code with application to nuclear fusion \cite{lupelli2015efit++}.
\item The ALF code performs analytics on data consumption log files
\item Spectre is a visualization tool providing near real time feedback on plasma and other operational conditions in fusion devices.
\item Savu provides a tomography reconstruction and processing pipeline \cite{wadeson2016savu}.
\end{itemize}
In order to support legacy and emerging data-analytics applications, we further develop widely used legacy HPC interfaces, such as MPI and HDF5, to exploit the SAGE platform.

\section{SAGE Platform Architecture} 
\label{sec-architecture} 
We next describe the high-level design of the SAGE system and then provide details on the SAGE platform architecture, based on the challenges and solutions outlined in the previous section. The fundamental requirements of the SAGE system justifies our top level architecture with an I/O hierarchy with in-storage compute capabilities, driven by an Object Storage infrastructure.  We will focus on the software stack in this paper.


\subsection{SAGE Hardware}
\begin{figure}
  \begin{center} 
  \includegraphics[width=0.7\linewidth]{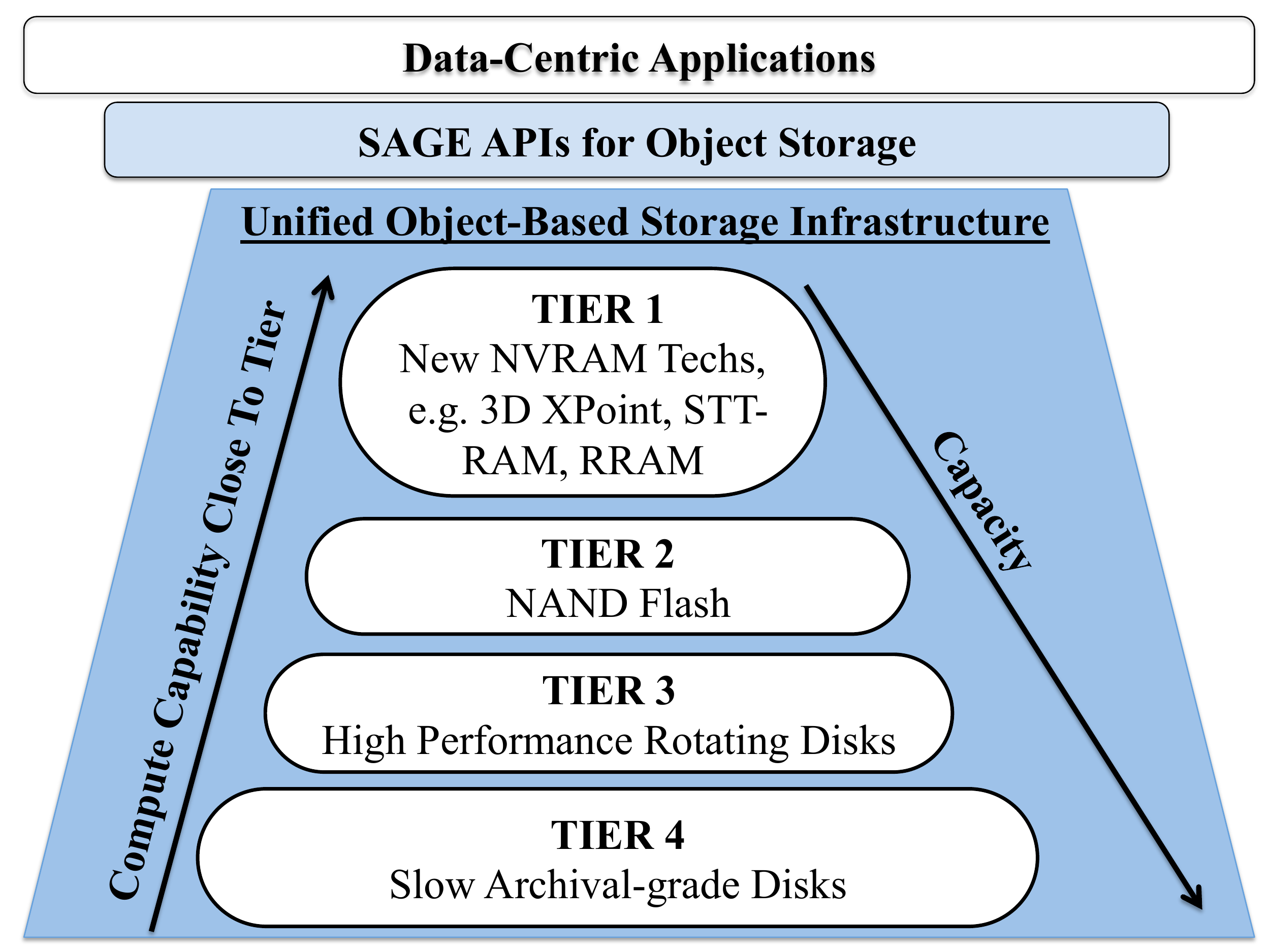}
  \caption{SAGE conceptual architecture.}
 \label{fig:Figure1}
 \end{center} 
\end{figure}

To drive discussion on the software stack, we briefly discuss the SAGE hardware. Figure  \ref{fig:Figure1} shows the conceptual architecture of the SAGE platform. 

The SAGE platform consists of multiple tiers of storage device technologies at the bottom of the stack, the \emph{Unified Object-Based Storage Infrastructure}. The system does not require any specific storage device technology type and accommodates  upcoming NVRAM, existing flash and disk tiers.  For the NVRAM tier, we are using Intel 3D XPoint technology~\cite{bourzac2017has} in our \emph{Tier 1}.  We will also use emulated NVDIMMs (Non-Volatile DIMMs) in Tier-1 because of the lack of NVDIMM availability in vendor roadmaps. We are using Flash based solid state drives in \emph{Tier 2}. Serial Attached SCSI high performance drives are contained in \emph{Tier-3} and archival grade, high capacity, slow disks ( based on Serial ATA and Shingled Magnetic Recording) are contained in \emph{Tier-4}. These tiers are all housed in standard form factor enclosures that provide their own compute capability, enabled by standard x86 embedded processing components, which are connected through an FDR infiniband network. Moving up the system stack, compute capability increases for faster and lower latency device tiers.

The base hardware platform which has been installed at the J\"ulich Supercomputing Center in 2017 and it will be described in a different report/paper. 

\subsection{SAGE Software Stack}
\begin{figure}
  \begin{center} 
  \includegraphics[width=0.7\linewidth]{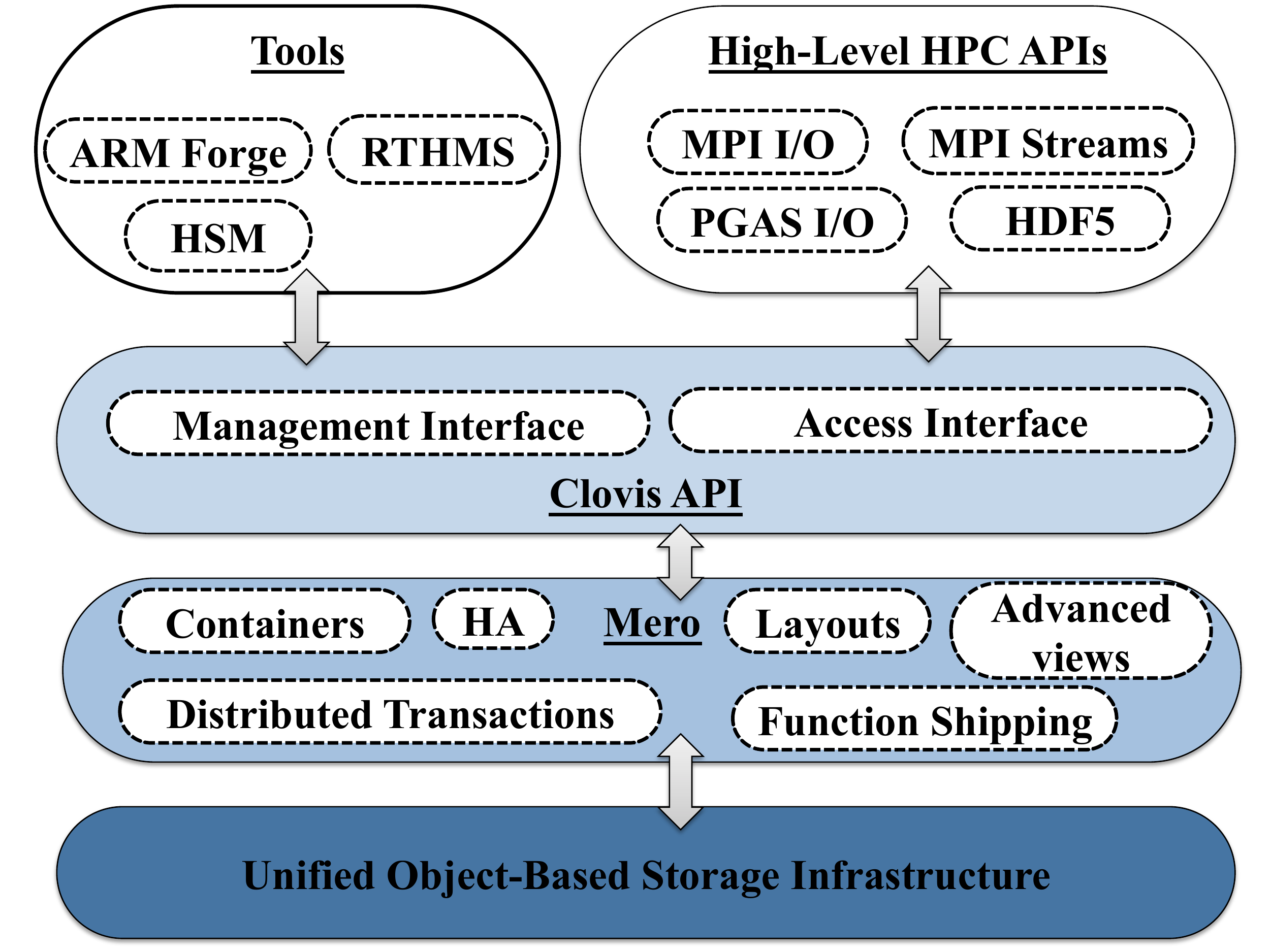} 
  \caption{SAGE System Stack.}
 \label{fig:Figure2}
 \end{center} 
\end{figure}

We have designed and developed a software stack capable of exploiting the SAGE Unified Object-Based Storage Infrastructure. A diagram of the SAGE software stack is shown in Figure \ref{fig:Figure2}. The SAGE software consists of three main layers: first, Mero is at the base of the software stack that provides the distributed object storage platform; second, the Clovis provides the higher-level layer with an interface to Mero; third, applications' APIs (MPI, HDF5, ...) and tools link applications with Clovis.

\subsubsection{Mero}
Mero is at the base of the SAGE software stack. Mero provides the Exascale-capable object storage infrastructure that drives the storage hardware \cite{danilov2016mero}. Mero provides the Exascale-capable object storage infrastructure that drives the storage hardware. Mero consists of a core that provides the basic object storage features that are typically available in many object stores. This consists of the ability to read/write objects with metadata for objects that can be defined in a Key Value store. The core also provides resource management of caches, locks, extents, etc on top of basic hardware and provides the basic infrastructure needed for hardware reliability such as distributed RAID enabled through Server Network Striping.

\textbf{Containers.} Containers are the basic way of grouping objects as per user definitions. Containers provide labelling of objects so as to provide a form of virtualisation of object name space. Containers can be based on performance(Eg: high performance containers for objects to be stored in higher tiers)  and data format descriptions(\emph{HDF5 containers}, \emph{NetCDF containers}, etc). There is in fact no limitation to the ways in which objects can be grouped into containers. Containers are also useful for performing one shot operations on objects such as shipping a function to a container.

\textbf{High Availability (HA) System.} Component reliability is not expected to increase in the coming years. Yet data has shown that the number of failures scales proportionally with respect to certain units, such as the amount of RAM, the number of cores, the number of NICs, etc. Making projections from existing systems ~\cite{failures} also indicates that we can expect several hardware failures per second at Exascale, in addition to software failures resulting in crashed nodes. To maintain service availability in the face of expected failures, the global state (or configuration) of the cluster may need to be modified, by means of a repair procedure.The HA subsystem for SAGE will perform automated repair activities within storage device tiers in response to failures. The subsystem monitors failure events (inputs) throughout the storage tiers. Then, on the basis of the collected events, the HA system decides whether to take action. The HA subsystem does not consider events in isolation but quantifies, over the recent history of the cluster, a quasi-ordered sets of events to determine which repair procedure (output) to engage, if any.

\textbf{Distributed Transaction Management.} Distributed transactions are groups of updates to the storage system that are guaranteed to be atomic with respect to failures. Transactions have now been recognized as a necessary component of a generic storage solution at scale ~\cite{FastForward}. On the other hand, traditional RDMS-style transactions are known not to scale. To address this problem, Mero separates transaction control proper from other issues, usually linked with it, such as concurrency control and isolation. The resulting transaction mechanism is scalable and efficient.

\textbf{Layouts.} A layout determines how a storage entity (an object, a key-value index, a container, etc.) is mapped to the available storage hardware and tiers. Layouts determine performance and fault-tolerant properties of storage entities. Each object has its own layout as described by the user of Mero. 
They determine how an object is distributed across multiple tiers of a deep storage hierarchy. Different types of layouts are possible. For example RAID layouts with different combinations of data and parity, compressed layouts, mirrored layouts, etc. Different portions of objects mapped to different tiers can have their own layout based on the property of the tier onto which they are mapped. 

\textbf{Function Shipping.} A system features horizontal scalability if its performance increases linearly (or quasi-linearly) with the number of components. An ideal storage cluster should scale horizontally: capacity and bandwidth should increase linearly with the number of nodes. However, the network may not scale proportionally, as is predicted for Exascale systems. Many computations can process data locally. That is, processing a portion of system input can be done independently of processing the rest of the input. In this case, it is not optimal to fetch raw data from a storage cluster into a compute cluster and process it there. In an Exascale situation, a large transfer overhead is predicted. Thus, computations should be distributed throughout the storage cluster and performed in place. Instead of moving the data to the computation, the computation moves to the data. The function-shipping component will provide the ability to run data-centric, distributed computations directly on the storage nodes where the data resides. Additionally, the computations offloaded to the storage cluster are designed to be resilient to errors. Well defined functions are offloaded from the use cases to storage through the API and invoked through simple Remote Procedure Call (RPC) mechanisms. 

\textbf{Advanced Views.} Considering the volume of objects in a distributed storage system at Exascale, it is quite wasteful to have different copies of objects to map to different types of data formats needed by different components of the application workflow, in case the same data is read. It is indeed quite desirable to have different ÒwindowsÓ into the same raw objects based on the applications using it. This is possible by manipulation of metadata associated with objects without copying the raw objects. This is termed as \emph{Advanced Views and Schemas} which we overlay on top of the Mero core. This will make it possible to have various views such as ÒS3 viewÓ, ÒHDF5 ViewÓ, ÒPOSIX viewÓ etc on top of the same set of objects.

\subsubsection{Clovis}
Clovis is the second SAGE software stack layer, sitting on top of Mero. Clovis is a rich, transactional storage API that can be used directly by user applications and can also be layered with traditional interfaces such as POSIX and RESTful APIs, much as libRados ~\cite{Librados} is the interface upon which the CephFS (POSIX), RadosGW (S3), and RBD (block device)~\cite{Ceph} interfaces are built. The Clovis I/O interface provides functions related to objects and indices for storing and retrieving data. Objects store traditional data. Indices store special data such as key-value pairs. Clovis object is an array of blocks. Blocks are of a power of two size bytes. This keeps the interface simpler and allows for efficient checks and translations between byte offsets and block indices \footnote{All block sizes occurring in practice (byte addressable storage, processor cachelines, block devices block sizes, etc.) are powers of 2}. They are of the same size for a particular object. The block size is selected when an object is created for the first time. Objects can be read from and written to at block level granularity. Objects can be deleted at the end of their lifetime. 

In the SAGE platform, Clovis consists of an access interface that provides access to objects (including through various gateway stacks), to specify containers and layouts for objects, shipped functions, and, transactional semantics. Clovis contains a management interface that accesses  telemetry records called Analysis and Diagnostics Data Base (ADDB) records on system performance that can be fed into external system data analysis tools. The Clovis management interface also contains an extension interface that will be used to extend the features and functionalities of Mero. 

\textbf{Clovis Access Interface.} Clovis consists of an access interface that provides access to objects (including through various gateway stacks), to specify containers and layouts for objects, and, transactional semantics. 

A Clovis index is a key-value store. An index stores records in some order. Records are key-value pairs with the constraint that keys are unique within an index. Clovis provides \textsf{GET}, \textsf{PUT}, \textsf{DEL} and \textsf{NEXT} operations on indices. The \textsf{GET} operation returns matching records from an index for a given set of keys. The \textsf{PUT} operation writes/rewrites a given set of records. The \textsf{DEL} operation deletes all matching records of an index for a given set of keys. The \textsf{NEXT} operation returns records corresponding to the set of next keys for a given set of keys.

\textbf{Clovis Management Interface.} Clovis contains a management interface that ADBB telemetry records on system performance that can be fed into external system data analysis tools. Clovis also contains an extension interface, as part of the management interface, that will be used to extend the features and functionalities of Mero. This interface is known as FDMI. Additional data management plug-ins can easily be built on top of the core through FDMI. Hierarchical storage management and information lifecycle management, file system integrity checking, data indexing, data compression are some examples of third-party plug-ins utilizing the API. 

\subsubsection{Tools}
A following are a set of tools for I/O profiling and optimized data movement across different SAGE platform tiers at the top of the SAGE software stack. 

\textbf{Data Analytics Tools.} Apache Flink, the data analytics tool employed in the SAGE project, will work on top of Clovis access interface through Flink connectors for Clovis. Using Flink enables the deployment of data analytics jobs on top of Mero. 

\textbf{Parallel File System Access.} Parallel file system access is the traditional method of accessing storage in HPC. Many of the SAGE use cases will need the support of POSIX compliant storage access. This access is provided through the pNFS gateway built on top of Clovis. However, pNFS will need some POSIX semantics (to abstract namespaces on top of Mero objects) to be developed by leveraging Mero's KVS. This abstraction is provided in SAGE. 

\textbf{HSM and Data Integrity.} HSM is used to control the movement of data in the SAGE hierarchies based on data usage. Advanced integrity checking overcomes some of the drawbacks of well known and widely used file system consistency checking schemes. 

\textbf{ARM Forge.} ADDB telemetry records from the Clovis management interface are directly fed to ARM Forge performance report tools that reports overall system performance for SAGE.

\textbf{RTHMS.} We designed and developed a tool, called RTHMS \cite{peng2017rthms}, that analyzes parallel applications and provides recommendations to the programmer about the data placement of memory objects on heterogeneous memory systems. Our tool only requires the application binary and the characteristics of each memory technology (e.g., memory latency and bandwidth) available in the system. 

\subsubsection{High-Level HPC Interfaces}
At the top of the software stack, we further develop widely-used HPC legacy APIs, such as MPI and HDF5, to exploit the SAGE architecture.

\textbf{PGAS I/O.} The goal of the Partitioned Global Address Space (PGAS) programming model is to provide processes with a global view of the memory and storage space during the execution of a parallel application. This is similar to what a Shared Memory model provides in a multithreaded local environment. In the PGAS approach, remote processes from different nodes can easily collaborate accessing memory addresses through load / store operations that do not necessarily belong to their own physical memory space. In SAGE, we propose an extension to the MPI one-sided communication model to support window allocations in storage: MPI storage windows~\cite{rivas2017mpi}. Our objective is to define a seamless extension to MPI to support current and future storage technologies without changing the MPI standard, allowing to target either files (i.e., for local and remote storage through a parallel file system) or alternatively address block devices directly (i.e., as in DRAM). We propose a novel use of MPI windows, a part of the MPI process memory that is exposed to other MPI remote processes, to simplify the programming interface and to support high-performance parallel I/O without requiring the use of MPI I/O. Files on storage devices appear to users as MPI windows (MPI storage windows) and seamlessly accessed through familiar \textsf{PUT} and \textsf{GET} operations. Details about the semantics of operations on MPI storage windows and the implementation are provided in Ref.~\cite{rivas2017mpi}.

In Section~\ref{sec-results}, we present the initial performance results of using MPI storage windows.

\textbf{MPI Streams for Post-Processing and Parallel I/O.} While PGAS I/O library addresses the challenge of heterogenous storage and memory, streams can be used to support function-shipping for post-processing and highly scalable parallel I/O. {\em Streams} are a continuous sequence of fine-grained data structures that move from a set of processes, called data {\em producers}, to another set of processes, called data {\em consumers}. These fine-grained data structures are often small in size and in a uniform format, called a {\em stream element}. A set of computations, such as post-processing and I/O operations, can be {\em attached} to a data stream. Stream elements in a stream are processed {\em online} such that they are discarded as soon as they are {\em consumed} by the attached computation. 

In particular, our work in SAGE focuses on {\em parallel streams}, where data producers and consumers are distributed among processes that require communication to move data. To achieve this, we have developed a stream library, called MPIStream, to support post-processing and parallel I/O operations on MPI consumer processes~\cite{peng2017mpi, peng2017preparing}. More details about MPI streams operation semantics and MPIStream implementation are provided in Ref.~\cite{peng2015data}. In Section~\ref{sec-results} we present the results of using MPI streams for post-processing and parallel I/O operations.

\textbf{HDF5.} Typically, data formats in HPC provide their own libraries to describe data structures and their relations (including I/O semantics). The HDF5 data format  needs to be supported in SAGE, and is layered directly on top of Clovis. The HDF5 will use the Virtual Object Layer Infrastructure within HDF5 (used to interface HDF5 with various object formats), to interface with Clovis.

\section{Results} 
\label{sec-results} 
We are starting to quantify the benefits of the individual features sets of the SAGE stack in the ongoing process of providing a holistic picture of the benefits of the SAGE architecture for Exascale. On that front, we present the performance results of \textbf{two} components of the high-level HPC interfaces for the SAGE platform: the PGAS I/O using MPI storage windows~\cite{rivas2017mpi} and MPI streams~\cite{peng2015data, peng2017mpi, peng2017preparing} for post-processing and efficient parallelI/O operations, exploiting the SAGE architecture. This aims to provide introductory results for SAGE stack components with results for other components targeted for future works. 
%
%
%

\subsection{PGAS I/O} 
In order to understand the potential performance constraints that the storage allocations introduce into the MPI one-sided communication model, we have carried out experiments with three benchmarks and applications~\cite{rivas2017mpi}:
\begin{itemize} 
\item {\bf STREAM} is a popular micro-benchmark to measure the sustainable memory bandwidth of a system~\cite{mccalpin1995survey}. As files are mapped into the MPI window, STREAM is a convenient benchmark to measure the access bandwidth to the MPI storage window and compare it with the bandwidth achieved when using MPI windows in memory. For this reason, we extend the MPI version of STREAM to support memory windows for each array allocation and instruct MPI to allocate the MPI windows on storage.
\item {\bf Distributed Hash Table (DHT)} mimics SAGE data-analytics applications that have random access in distributed data structures. The source code is based on the implementation of a DHT presented in Ref.~\cite{gerstenberger2014enabling}. 
\item {\bf HACC} is a physics particle-based code, simulating the trajectories of trillions of particles. For our tests, we use the HACC I/O kernel to mimic the check-pointing and restart functionalities in the SAGE iPIC3D application~\cite{markidis2010multi}. We extend the kernel to perform check-pointing and restart using MPI storage windows to compare with the existing MPI I/O implementation.
\end{itemize}

As the SAGE system has been only recently installed, we tested the SAGE software components on other workstations and supercomputers available within the SAGE consortium. To carry out different experiments and obtain preliminary results about the SAGE software stack, we use two testbeds, specified as follows:
\begin{itemize} 
\item Blackdog is a workstation with eight-core Xeon E5-2609v2 processor running at 2.5GHz. The workstation is equipped with a total of 72GB DRAM. The storage consists of two 4TB HDD (WDC WD4000F9YZ / non-RAID) and a 250GB SSD (Samsung 850 EVO). The OS is Ubuntu Server 16.04 with Kernel 4.4.0-62-generic. Compilation uses gcc v5.4.0 and MPICH v3.2.
\item Tegner is a supercomputer at KTH. We use up to six compute nodes that are equipped with Haswell E5-2690v3 processor running at 2.6GHz. Each node has two sockets with 12 cores and a total of 512GB DRAM. The storage employs a Lustre parallel file system. The OS is CentOS v7.3.1611 with Kernel 3.10.0- 514.6.1.el7.x86\_64. Compilation uses gcc v6.2.0 and Intel MPI v5.1.3.
\end{itemize} 

\begin{figure}
  \begin{center} 
  \includegraphics[width=0.7\linewidth]{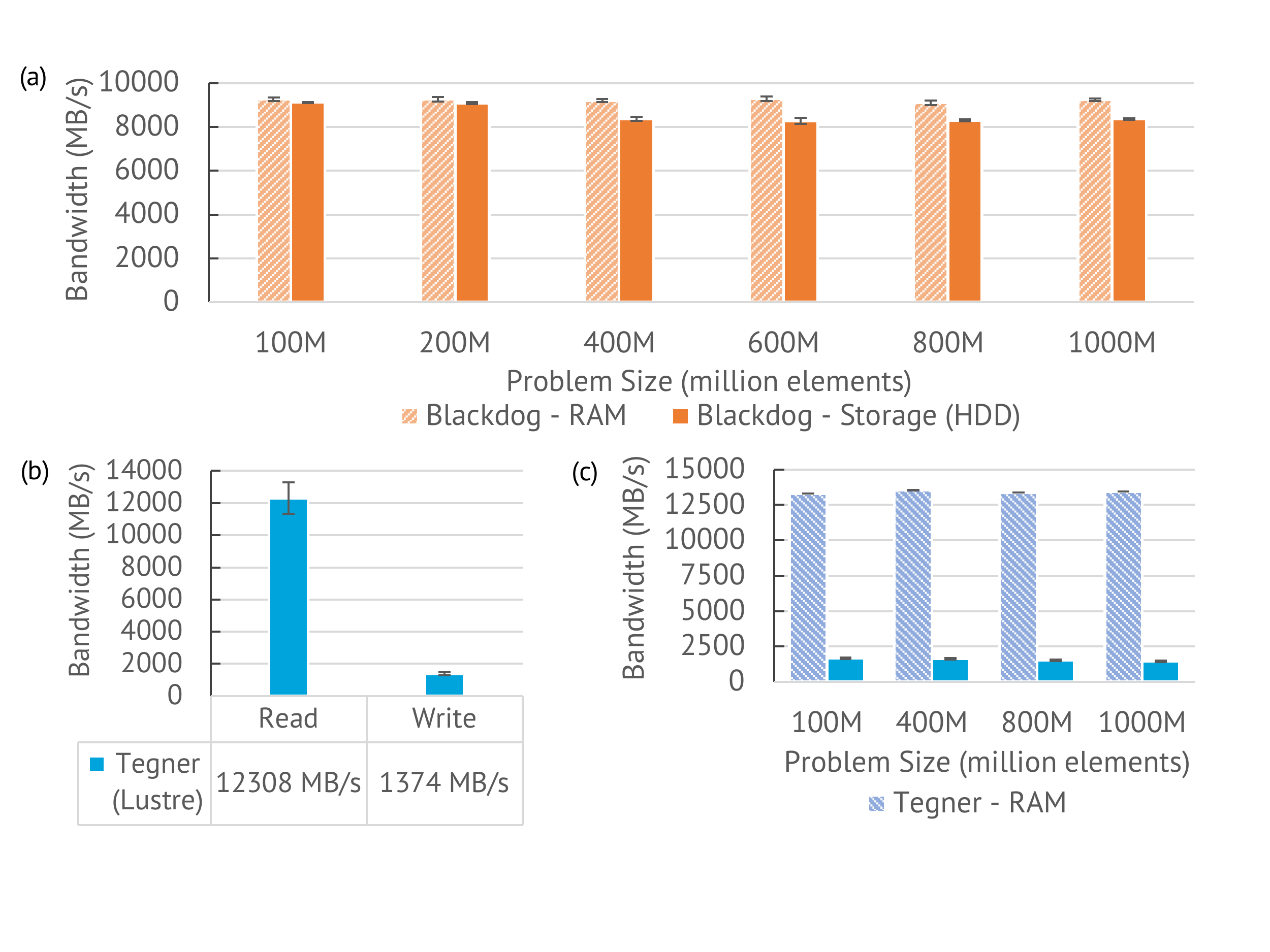}
  \caption{STREAM benchmark performance using MPI windows on storage and DRAM on two different systems.}
 \label{fig:FigureSTREAM}
 \end{center} 
\end{figure}

Figures~\ref{fig:FigureSTREAM}(a) and~\ref{fig:FigureSTREAM}(c) present the bandwidth of the modified STREAM benchmark on Blackdog and Tegner respectively with MPI windows and MPI storage windows. The x-axis shows the problem size in millions of elements per array, while the y-axis presents the measured bandwidth. Figure \ref{fig:FigureSTREAM}(b) shows the asymmetric bandwidth for read / write on Tegner, using Lustre as the underlying parallel file system and the copy kernel of STREAM.

Figure~\ref{fig:FigureSTREAM}(a) shows that only approximately a 10\% degradation is observed on Blackdog  for the largest case of 1000 million elements per array. Analyzing Figure~\ref{fig:FigureSTREAM}(c), we observe that the MPI storage window performance on Tegner degrades the bandwidth by 90\% compared to the MPI window allocation on memory. In Figure~\ref{fig:FigureSTREAM}(b) we can observe that read operations can reach as high as 12,308MB/s and write operations only reach 1,374MB/s. This explains why the STREAM benchmark, which constantly enforces write operations between the three arrays, performed considerably worse and did not take advantage of the read bandwidth.

The second application we test the MPI storage windows is the Distributed Hash Table (DHT). In the DHT application, each MPI process handles a part of the DHT, named Local Volume. These volumes have multiple buckets to store elements individually. The processes also maintain an overflow heap to store elements in case of collisions. For instance, for 8 processes and a conflict overflow of 4 per element, 100 million elements per local volume means a capacity of 500 million elements per MPI process and a global DHT size of 4000 million elements (800 million excluding the overflow heap). The local volume and the overflow heap are allocated as MPI windows on each process, so that updates to the DHT are handled using MPI one-sided operations. In this way, each MPI process can put or get values and resolve conflicts asynchronously on any of the exposed local volumes.

\begin{figure}
  \begin{center} 
  \includegraphics[width=0.7\linewidth]{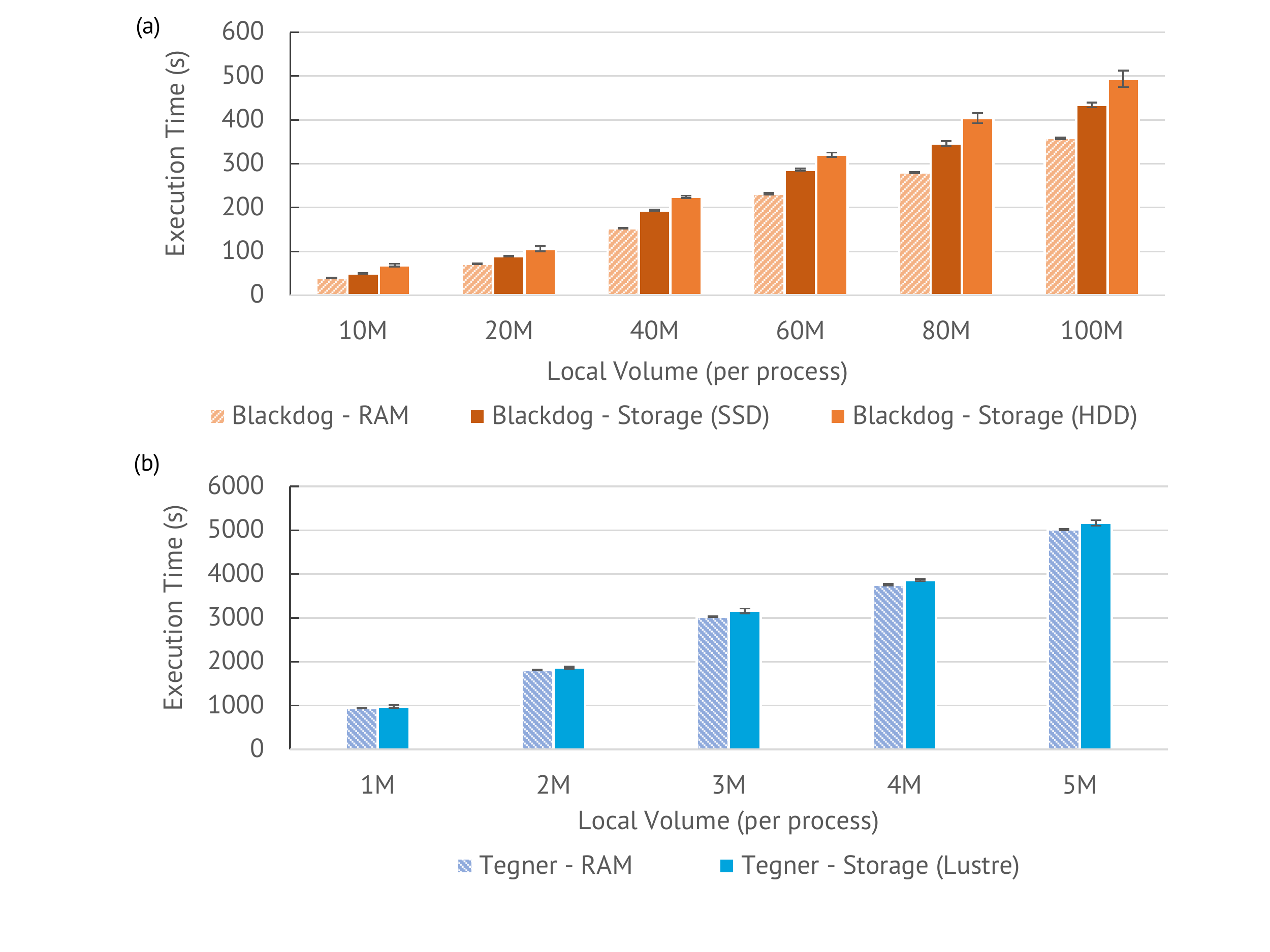}
  \caption{DHT application performance using MPI windows on storage and DRAM on two different systems shows that performance penalty of using storage instead of memory is relatively small.}
 \label{fig:FigureDHT}
 \end{center} 
\end{figure}

Figures~\ref{fig:FigureDHT}(a) and~\ref{fig:FigureDHT}(b) present the execution time of the DHT on Blackdog and Tegner, respectively, using MPI windows and MPI storage windows. The x-axis represents the Local Volume per process in millions of elements, while the y-axis shows the average execution time. The execution on Blackdog is using eight MPI processes, while the execution on Tegner is using 96 MPI processes (4 nodes).

The DHT results on Blackdog in Figure~\ref{fig:FigureDHT}(a) demonstrate that the overall overhead of using MPI storage windows with conventional hard disks is 34\% compared to the memory-based approach. However, one of the advantages that the SAGE system will feature is its multi-tier storage capabilities, containing Non-Volatile RAM (NVRAM) together with other storage technologies. Consequently, we evaluated the performance once again by using a faster storage device, such as a solid-state drive (SSD). The performance clearly improves in comparison by decreasing the overhead to approximately 20\% on average. In the case of Tegner, Figure~\ref{fig:FigureDHT}(b), using MPI storage windows barely affects the performance with only 2\% degradation on average when compared to MPI memory windows.

Finally, we evaluate the MPI storage window performance with HACC I/O kernel (strong scaling). We use 100 million particles in all the tests, while increasing the number of processes in use. We ensure synchronization both during check-pointing and restart for fair comparison with MPI I/O. 

\begin{figure}
  \begin{center} 
  \includegraphics[width=0.7\linewidth]{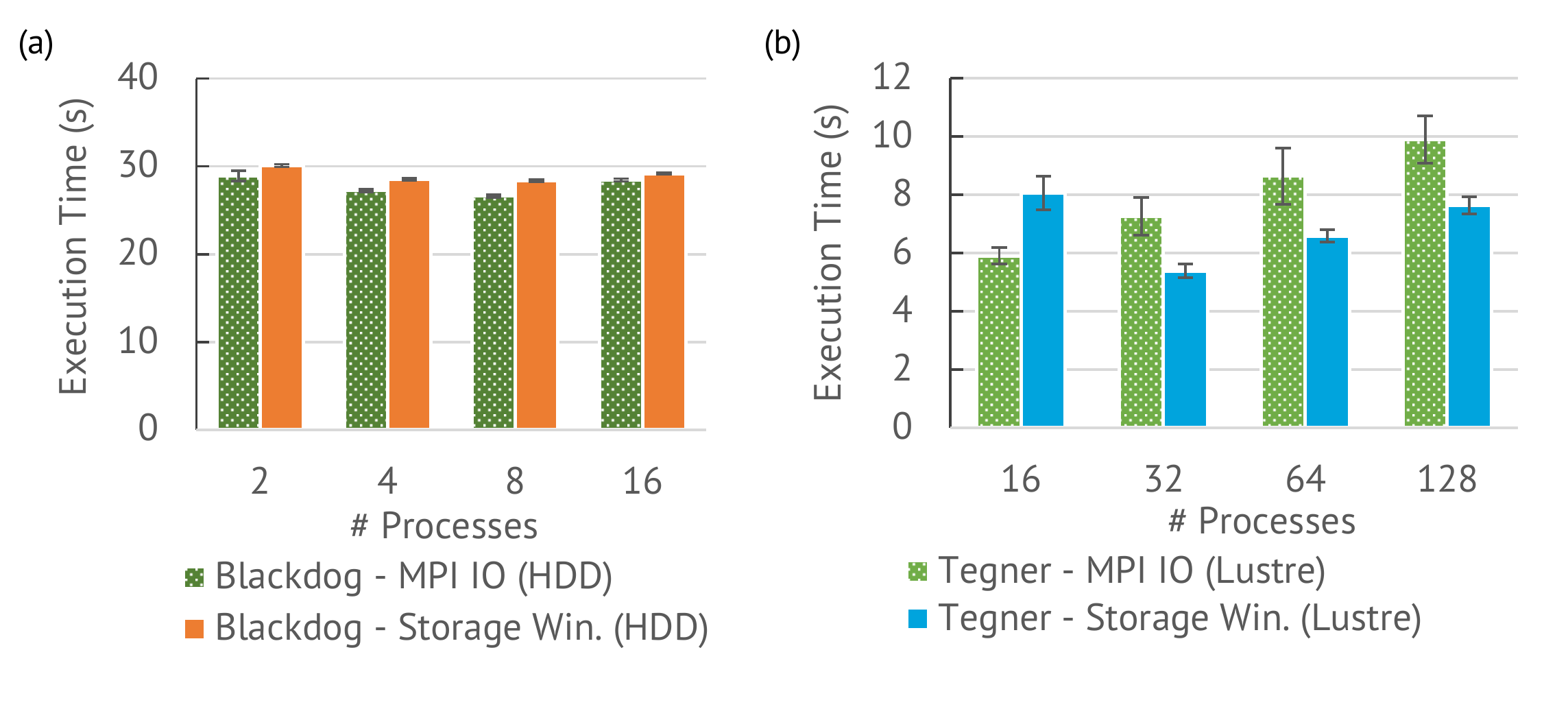}
  \caption{Execution time of the HACC I/O kernel, mimicking iPIC3D checkpointing / restart, running on Blackdog and Tegner using MPI I/O and MPI storage windows. MPI storage windows provide better scalability when compared to MPI I/O on a larger number of processes.}
 \label{fig:FigureHACC}
 \end{center} 
\end{figure}

Figure~\ref{fig:FigureHACC} shows the average execution time on Blackdog and Tegner when varying the number of processes. The results show that MPI storage windows provide about 32\% improvement on average on Tegner when compared to the MPI I/O implementation when the process count increases. On the other hand, the performance of the two approaches are similar on Blackdog, with MPI I/O performing slightly better (4\% on average). While the performance difference between the two approaches is relatively small, this result indicates that the MPI storage windows provide better scalability compared to MPI I/O on a larger number of processes.  

In SAGE, we showed that MPI storage windows provide a high-performance I/O mechanism. The bandwidth to MPI storage windows is only a fraction smaller than the bandwidth to MPI windows in memory. Parallel I/O with MPI storage windows is also faster than traditional parallel I/O solutions, such as MPI I/O, when the number of processes is large. High-Performance parallel I/O is achieved by the use of memory-mapped file I/O within the MPI storage windows. In fact, the OS page cache and buffering of the parallel file system act as automatic caches for read and write operations on storage: in the same way that programmers do not necessarily handle explicit data movement in the processor caches, here programmers do not need to handle virtual memory management or buffering on the file system, i.e., the OS and underlying file systems provide this functionality.

\subsection{MPI Streams for Post-processing and Optimized I/O} 
In SAGE, we developed MPI streams to support offloading of post-processing and parallel I/O to a small number of processes to enable high-performance I/O. As future tasks, we plan to use Clovis function shipping capabilities to perform operations on streams directly on storage.  

\begin{figure}
  \begin{center} 
  \includegraphics[width=0.7\linewidth]{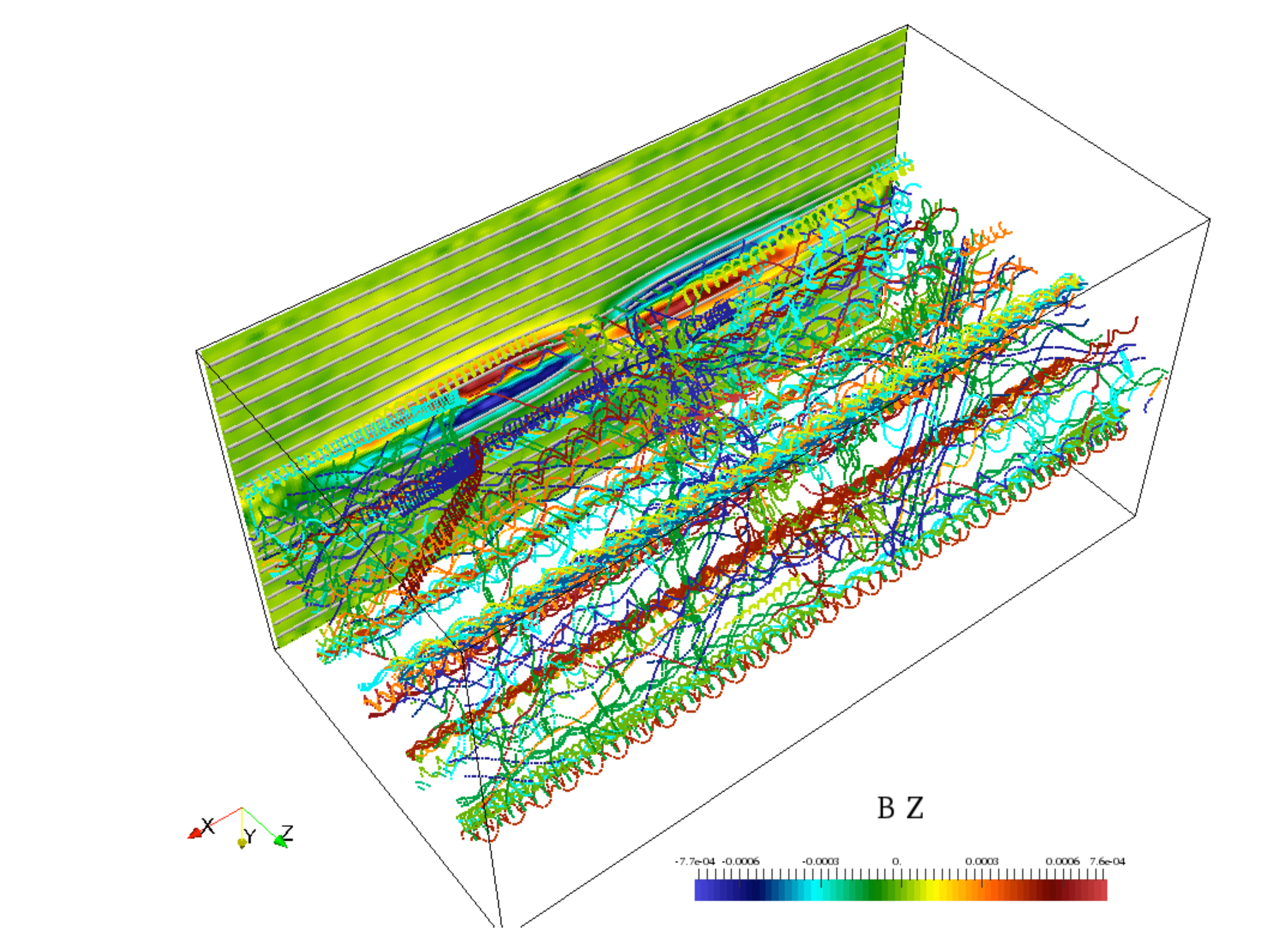}
  \caption{MPI streams allows iPIC3D to offload particle visualization and parallel I/O to a smaller number of MPI processes and continue with the simulation. As next step in SAGE, Clovis function-shipping capabilities can be used for post-processing on storage.}
   \label{fig:FigureParticles}
 \end{center} 
\end{figure}

To test the performance of MPI Streams for I/O, we use one of the SAGE applications, iPIC3D~\cite{markidis2010multi}. The current production version of the iPIC3D code uses the MPI collective I/O for saving snapshots of relevant quantities, such particle positions and velocities, to disk. In this use case, we use MPI Streams to offload post-processing to separate a program that performs I/O and visualization at runtime. In this way, MPI processes that carry out the simulation are isolated from the frequent and expensive I/O operations~\cite{peng2017mpi}. We use the MPI Streams library to decouple the processes from I/O operations by streaming out the particle data to the I/O program so that simulations can proceed without carrying out I/O operation. Concurrently, the I/O and visualization program continues processing the received particle data. A visualization of high energy particles trajectories with Paraview application is presented in Figure \ref{fig:FigureParticles}.

In this case, particle data is streamed out to the I/O and Paraview program at a frequency as high as each time step. Thus, no data is lost for these particles of interest and their motion can be tracked accurately. In this case, the iPIC3D code is the data producer and the I/O and post-processing code is the data consumer. The stream element is the basic unit of the communication between these two programs. It is defined as the structure of a single particle that consist of eight scalar values: particle position \textsf{(x,y,z)}, particle velocity \textsf{(u,v,w)}, particle charge \textsf{q} and an identifier \textsf{ID}. For tracking high energy particles, only those particles with energy exceeding certain thresholds are streamed out. It is unpredictable when and which particles will reach high energies. Thus, a particle is streamed out during particle mover, where the location and velocity of each particle is calculated. Once a particle reaches high energies, it is continuously tracked in the remaining of the simulation. The I/O and visualization program continues receiving particle streams from the simulation at runtime and processing them to prepare data in file formats, such as VTK, that can be visualized on-the-fly by the Paraview application. The I/O and visualization program can flush data to the file system at a user-defined frequency. With sufficiently high frequency, the user can visualize the real-time motion of particles during simulation. 

We perform the performance tests of MPI streams for I/O on the KTH Beskow supercomputer. Beskow is a Cray XC40 supercomputer with Intel Haswell processors and Cray Aries interconnect network with Dragonfly topology. The supercomputer has a total of 1,676 compute nodes of 32 cores divided between two sockets. The operating system is Cray Linux, and the applications are compiled with the Cray C compiler version 5.2.40 with optimization flag -O3 and the Cray MPICH2 library version 7.0.4.

\begin{figure}
  \begin{center} 
  \includegraphics[width=0.7\linewidth]{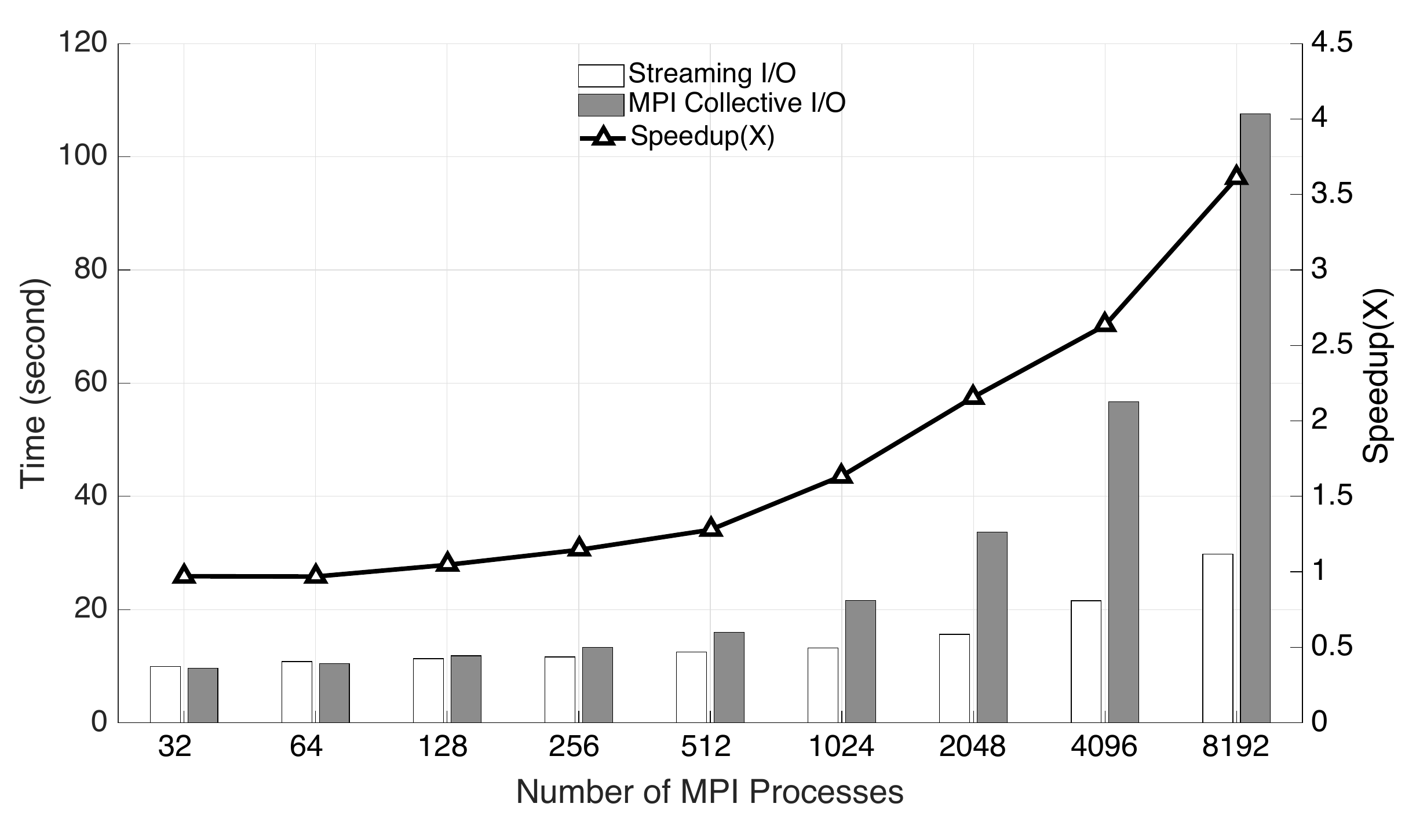}
  \caption{MPI streams allows iPIC3D to offload parallel I/O to a small number of MPI processes achieving considerable performance gain. Data can be streamed to Clovis clients to perform I/O on the object storage.}
 \label{fig:scalingParticles}
 \end{center} 
\end{figure}

Offloading I/O operations to a smaller number of I/O processes can reduce the communication time. The performance gain from the streaming model increases as the scale of the system increases. This is visible from Figure~\ref{fig:scalingParticles} that presents the scaling test results comparing the particle visualization using MPI collective I/O (in grey bars) and the streaming model for decoupling I/O from simulation (in white bars). The streaming I/O uses one visualization process for every 15 simulation process.

The simulation runs for 100 time steps and the total execution time is shown in y-axis. The improvement is calculated by dividing the execution time in MPI collective I/O over the execution time in the streaming model. The improvement results are shown in the solid line against the secondary y-axis. On small number of processes, the two approaches show comparable performance. Starting from 256 processes, the streaming model demonstrates a steady improvement that continues increasing to 3.6X speedup on 8,192 processes. The observed results are in line with the fact that I/O operations on supercomputers are often a major performance bottleneck and shows that performance gain by using MPIStream, developed in SAGE, for I/O can be considerable. Additional details about MPI stream performance for optimizing parallel I/O and post-processing are reported in Refs.~\cite{peng2017mpi,peng2017preparing}. 

As future tasks, we plan to stream data to Clovis clients and perform I/O on the object storage.

\section{Related Work} 
\label{sec-relwork} 
To the best of our knowledge SAGE is the first HPC-enabled storage system to implement new NVRAM tiers, Flash and disk drive tiers as part of a single unified storage system. 
The SAGE Architecture progress the state of the art from Blue Gene Active Storage~\cite{fitch2010blue} and Dash~\cite{he2010dash}, which use flash for data staging. SAGE also progress the state of the art from Burst Buffer technologies as discussed earlier. 

We note that Data Elevator~\cite{Dataelevator} addresses some specific aspects of SAGE (primarily HSM)  of moving data transparently between multiple tiers in a hierarchical storage system. 

SAGE highly simplifies storage compared to what was developed in the FastForward Project~\cite{FastForward} and develops a solution for deep I/O hierarchies, including NVRAM technologies. Further, the FastForward solution is evolutionary as it tries to make use of an existing storage solution, namely, Lustre~\cite{schwan2003lustre} used for the last 20 years or so, that was really designed for the previous era, when use cases and architectural assumptions were different. SAGE and Mero are the product of a complete redesign in consideration of the new requirements arising out of the Extreme scale computing community. 

Mero, the object store in SAGE, extends the state of the art in existing object storage software, Ceph~\cite{weil2006ceph} and Open Stack Swift ~\cite{swift} by building Exascale components required for extreme scale computing. Fundamentally Ceph and Openstack swift are designed for cloud storage, whereas Mero is built to meet the needs of the extreme scale computing community. 

%
%

\section{Conclusions and Future Work}
\label{sec-conclusions}
We aimed at designing and implementing an I/O system capable of supporting I/O workloads of Exascale supercomputers. The SAGE platform, recently installed at Julich Computing center, supports a multi-tiered I/O hierarchy and associated intelligent management software, to provide a demonstrable path towards Exascale. The SAGE software stack consists of three main software layers: the Seagate Mero object-storage, the Clovis API with tools and high-level interfaces, such as MPI, on the top of the software stack. We presented the performance results of the first implementation of high level HPC interfaces for SAGE. The current work focuses on porting the higher level interfaces and tools to the SAGE system.

Part of ongoing work is  the focus on the performance characterization of various new NVRAM device technologies appearing in the time frame of the SAGE project. This is also looking at lower level software  and Operating System(OS) infrastructure  requirements to exploit these  new devices types,  below Mero in the SAGE stack. We clearly recognize that various NVRAM technologies have their own performance characteristics and limitations. New NVRAM technologies can be part of the SAGE hardware tiers based on where they ultimately are on the performance and capacity curve. The SAGE stack and Mero indeed is designed to be agnostic of storage device types as long as adaptations are in place within the  OS. 

The next steps will be to quantify the benefits of the various features of the SAGE stack on the SAGE prototype system currently installed at Juelich Supercomputing Center, with focus on providing results for the remaining SAGE components and the SAGE architecture as a whole. As a part of this external organisations outside of the SAGE consortium ( eg: from Climate and Weather, Astronomy, etc) will soon be granted access to study how their codes and worlflows can exploit the features of the SAGE platform.  We will then look at extrapolation studies of the benefits of the various SAGE features at Exascale through analytical and simulation models. These will be discussed separately. Porting of the SAGE stack across other sites and extensions of the SAGE prototype is also planned. We are targeting SAGE work to be a part of European Extreme Scale Demonstrators~\cite{esd} which will be pre-exascale prototypes. 
%
%

\section*{Acknowledgements}
The authors acknowledge that  the SAGE work is being performed by a consortium of members consisting of Seagate(UK), Bull ATOS(France), ARM(UK), KTH(Sweden), STFC(UK), CCFE(UK), Diamond(UK), DFKI(Germany), Forchungszentrum Juelich(Germany) and CEA(France) - which is being represented by the authors in this paper. 

Funding for the work is received  from the European Commission H2020 program, Grant Agreement No. 671500(SAGE). 

\section*{References}


\end{document}